\documentclass[sigplan,nonacm]{acmart}

\usepackage{amsmath}
\usepackage{algorithm}
\usepackage{algpseudocode}
\usepackage[table,xcdraw]{xcolor}
\usepackage{booktabs}
\usepackage{array}
\usepackage{stfloats}

\usepackage{booktabs}
\usepackage{multirow}
\usepackage{booktabs}
\usepackage{tabularx}
\usepackage{makecell}
\AtBeginDocument{%
  }

\begin{document}


\title{SeqMoE: Toward Full-Load Performance via Predictive and Graph-Compatible MoE Offloading}


\author{Zihan Wang}
\affiliation{%
  \institution{University of Science and Technology of China}
  \city{Hefei}
  \country{China}}
\email{wangzh196@mail.ustc.edu.cn}

\author{Yuqi Wang}
\affiliation{%
  \institution{University of Science and Technology of China}
  \city{Hefei}
  \country{China}}
\email{xiaoqi17@mail.ustc.edu.cn}

\author{Lei Gong}
\affiliation{%
  \institution{University of Science and Technology of China}
  \city{Hefei}
  \country{China}}
\email{leigong0203@ustc.edu.cn}

\author{Cheng Tang}
\affiliation{%
  \institution{University of Science and Technology of China}
  \city{Hefei}
  \country{China}}
\email{sisyphustc@mail.ustc.edu.cn}

\author{Wenqi Lou}
\affiliation{%
  \institution{Suzhou Institute for Advanced Research, University of Science and Technology of China}
  \city{Suzhou}
  \country{China}}
\email{louwenqi@ustc.edu.cn}

\author{Teng Wang}
\affiliation{%
  \institution{Suzhou Institute for Advanced Research, University of Science and Technology of China}
  \city{Suzhou}
  \country{China}}
\email{wangt635@ustc.edu.cn}

\author{Chao Wang}
\affiliation{%
  \institution{University of Science and Technology of China}
  \city{Hefei}
  \country{China}}
\email{cswang@ustc.edu.cn}

\author{Xuehai Zhou}
\affiliation{%
  \institution{University of Science and Technology of China}
  \city{Hefei}
  \country{China}}
\email{xhzhou@ustc.edu.cn}







\begin{abstract}
Mixture-of-Experts (MoE) creates a structural advantage for offloading: only a small fraction of activated experts need to reside in device memory, and if they can be loaded in time for computation, offloading can in principle approach full-load performance, where all model weights reside in device memory. Yet translating MoE’s structural advantage into practical offloading gains remains challenging. We propose SeqMoE to bridge this gap. To maximize expert hits, we build predictive memory management: (i) Sequence-to-sequence prediction. We are the first to recast expert activation prediction as sequence modeling, enabling accurate multi-step, multi-layer forecasts that provide a long and reliable window for downstream decisions. (ii) Joint prefetch scheduling. We formulate prefetch scheduling as Job Sequencing with Deadlines to maximize expected expert hits and improve bandwidth efficiency. (iii) Forecast-driven caching. Leveraging the recursive nature of sequence modeling, we introduce a probabilistic Belady policy for future-aware eviction. To eliminate execution bottleneck, we develop (iv) Graph-compatible offloading runtime. We derive general runtime principles encompassing compute-transparent expert placement and synchronization-free orchestration disciplines for end-to-end graph capture. With 45\% expert residency, SeqMoE averages a 96.97\% hit rate and 80.22\% of full-load performance, advancing the state of the art in MoE offloading.
\end{abstract}



\maketitle

\section{Introduction}
Growing demands for personalization, privacy, embodied intelligence, and autonomous driving are pushing large language models (LLMs) toward the edge~\cite{appleedge}. However, limited edge-device memory hinders large-model deployment. Offloading inference addresses this mismatch and has emerged as a research focus~\cite{offloadingsurvey1, offloadingsurvey2}. Its core principle is to treat device memory (e.g., GPU HBM) as a cache that holds only a subset of model weights, while relegating the rest to larger next-level memory (e.g., CPU DRAM). By extending usable capacity beyond device memory, offloading brings the intelligence of larger models to edge applications.

\begin{figure}[t]
  \centering
  \includegraphics[width=\linewidth]{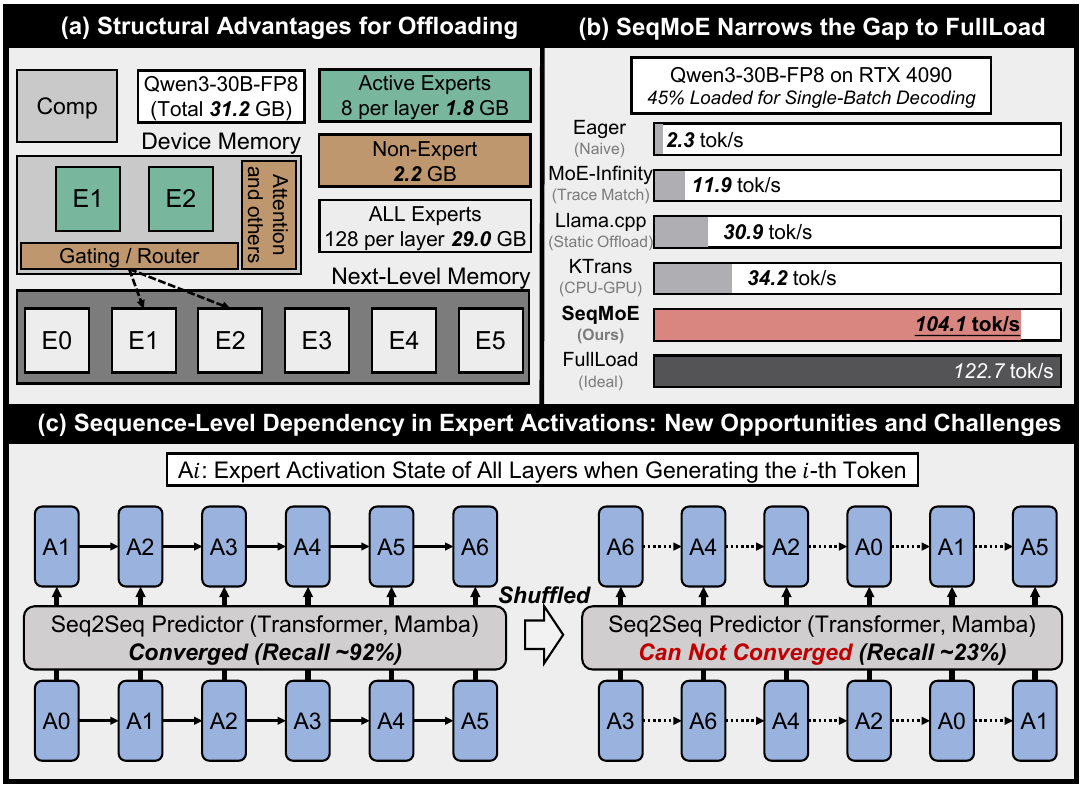}
  \caption{Realizing MoE’s Offloading Advantage.}
  \Description{...}
  \label{fig:intro}
\end{figure}

Mixture-of-Experts (MoE), a standard architecture for modern LLMs, is particularly amenable to offloading. As shown in Fig.~\ref{fig:intro} (a), its structural advantage stems from sparse scaling: MoE layers dominate the model weights (93\%), yet only a small subset of experts (6\%) are activated at each inference step. This allows non-MoE layers to remain resident in device memory, while all experts reside in next-level memory and only activated ones need to be dynamically loaded into the device cache. With such dynamic sparsity, \textbf{if all activated experts can be loaded in time for computation, offloading can in principle approach FullLoad performance, where all model weights reside in device memory. This defines the ideal target for offloading inference}.

However, translating MoE’s structural advantage into practical offloading gains remains challenging. The key bottleneck is load-on-demand stalls caused by expert misses, which expose expensive memory I/O on the critical path. As shown in Fig.~\ref{fig:intro} (b), eager offloading achieves only about 2\% of FullLoad performance. Despite extensive efforts to mitigate this bottleneck, existing systems remain far from FullLoad. In contrast, SeqMoE not only substantially outperforms existing MoE offloading systems, but also can approach FullLoad performance, unleashing MoE’s structural advantage for high-performance, low-memory inference. \textbf{To achieve this, we take maximizing expert hits as the primary objective and build predictive memory management that jointly optimizes early prediction, prefetch scheduling, and cache residency. As load-on-demand stalls recede, we further develop a graph-compatible offloading runtime to eliminate the emerging execution bottleneck}. 

\textit{Activation sequence modeling achieves long overlap windows and high accuracy simultaneously}. Expert activation is late-bound, so early prediction is essential for downstream decisions. Mainstream methods stay within a single autoregressive step, predicting deeper-layer expert choices from shallow-layer states. Their accuracy decays rapidly with prediction distance, leaving a prefetch window of at most three layers~\cite{promoe,moeapex,moeinfinity,tamingmoe}. A few recent works attempt cross-step prediction, but rely on per-layer conditional activation statistics between adjacent tokens. Such a memoryless, first-order estimate is too inaccurate for prefetch decisions~\cite{stmoe,patternsmoe}. In contrast, as shown in Fig.~\ref{fig:intro} (c), we encode the activation state of the full layer stack per generation step and train a Transformer~\cite{transformer} probe to perform next-step prediction. Convergence holds on the ordered sequence and disappears under step shuffling, indicating that expert activation carries learnable sequence-level dependencies~\cite{seq2seq} (Section~\ref{sec:Expert Activation Sequence Modeling}). Therefore, we recast expert activation prediction as sequence modeling, extending the overlap window across autoregressive steps at high accuracy. 

\textit{Activation sequence prediction extends overlap windows, but complicates prefetch scheduling}. A longer overlap window relaxes the horizon constraint but not bandwidth: a slow link such as PCIe sustains only a fraction of prefetch tasks, so the prefetcher must decide which experts deserve bandwidth. Candidates differ along two axes. Prediction confidence is an objective: it quantifies the transfer's expected contribution to expert hits. Deadline is a constraint: it temporally couples tasks into bandwidth contention. Existing prefetchers overlook the joint scheduling structure and conflate both into independent per-task scores~\cite{promoe, moeinfinity}, leading to suboptimal selection (Section~\ref{Joint Structure in Prefetch Scheduling}). This waste grows as longer overlap windows deepen task queue and enlarge scheduling space. Instead, we formulate prefetching as Job Sequencing with Deadlines~\cite{jobsequence} to improve bandwidth efficiency.

\textit{Activation sequence forecasting transforms history-statistics caching into future-aware eviction}. Effective caching retains reusable experts, avoiding redundant transfers and reducing both prefetch bandwidth pressure and expert misses. Existing policies typically adapt classical eviction algorithms such as LRU and LFU, collapsing historical activation states into coarse statistics to extrapolate reuse~\cite{freetoken, moeapex}. We compare them against variants of the optimal Belady’s MIN~\cite{beladymin} and show that exploiting future activations yields consistently better eviction decisions (Section~\ref{sec:Future-Aware Cache Eviction}). Fortunately, the recursive nature of sequence modeling makes future-aware eviction practical: rather than relying on lossy historical summaries, we roll the historical sequence forward into multi-step forecasts to guide cache eviction. 

\textit{As expert misses recede, offloading runtime emerges as the new execution bottleneck}. An offloading runtime is essential to coordinate the offloading pipeline with LLM inference. However, existing runtimes introduce substantial coordination overhead that prevents execution from achieving full-load efficiency. The fundamental limitation is their incompatibility with graph-based execution mechanisms~\cite{graphexe}, such as CUDA Graphs~\cite{nvidiagraph}, that reduce dispatch overhead by replaying pre-captured operations (Section~\ref{sec:End-to-End Graph Execution}). First, dynamic expert placement in the cache hinders fused MoE kernels~\cite{tritonfusedmoe, deepgemm}, a key enabler of graph-based MoE execution. Second, prefetching and caching rely heavily on host-device synchronization, which fragment graph execution and diminish its benefits. FreeToken~\cite{freetoken}\footnote{A recent concurrent work relying solely on LRU for expert hits.} explores graph-based offloading but is misaligned with predictive management, which is critical to high expert-hit rates. In contrast, we derive general runtime principles that coordinate predictive memory management with LLM inference for end-to-end graph capture.

Building on these insights, we present SeqMoE, an MoE offloading system that delivers low-memory inference approaching full-load performance. To maximize expert hits, we build predictive memory management: (i) \textit{Sequence-to-Sequence (seq2seq) expert activation predictor}. We are the first to recast expert activation prediction as sequence modeling, capturing routing dependencies across the full layer stack and generation trajectory. We adopt Mamba2~\cite{mamba2}, a lightweight and efficient sequence model, as the predictor backbone. During inference, the predictor continuously forecasts multi-step, multi-layer activations, providing long and reliable windows for downstream decisions. (ii) \textit{Joint prefetch scheduling}. We formulate prefetch scheduling as a matroid optimization problem~\cite{matroids} to maximize expected expert hits under deadline constraints, thereby improving bandwidth efficiency. (iii) \textit{Forecast-driven caching}. Leveraging the predicted sequence of future activations, we introduce probabilistic Belady eviction, which prioritizes experts by their likely near-term reuse, effectively retaining reusable experts. To eliminate execution bottleneck, we develop (iv) \textit{Graph-compatible offloading runtime}. We derive general principles encompassing compute-transparent expert placement for fused MoE kernels and synchronization-free offloading orchestration disciplines. They coordinate predictive memory management with LLM inference for graph-based execution.


We evaluate SeqMoE across diverse datasets on Qwen3-30B-A3B, Qwen3.6-35B-A3B, GPT-OSS-120B, and DeepSeek-V4-Flash against llama.cpp~\cite{llamacpp}, MoE-Infinity~\cite{moeinfinity}, KTransformers~\cite{ktransformers}, and FreeToken~\cite{freetoken}. At 25\% and 45\% expert residency, SeqMoE averages 91.72\% and 96.97\% hit rates, versus 74.24\% and 88.50\% for the strongest baseline. At 45\% residency, it averages 80.22\% of full-load performance, versus 51.50\% for FreeToken and at most 28.03\% for other baselines.
\begin{itemize}
    \item We are the first to formulate expert activation prediction as sequence modeling, enabling long overlap windows with high accuracy.
    \item We propose predictive memory management for MoE offloading, jointly optimizing early prediction, prefetching, and cache residency to maximize expert hit rates.
    \item We propose a graph-compatible runtime that enables graph-based execution of MoE offloading.
    \item Results show that SeqMoE approaches full-load performance and advances the state of the art in MoE offloading inference.
\end{itemize}

\section{Background}
\subsection{MoE's Structural Advantages for Offloading}
Mixture-of-Experts (MoE) has become a standard architecture for modern LLMs, employing a router to dynamically select active experts~\cite{mixtralmoe}. Consider an MoE model with $L$ layers, hidden dimension $d$, and $e$ experts per layer, of which $k$ experts are activated per token. A request comprises $I$ token steps spanning both prefill and decoding, with batch size $b$. At step $i\in [0,I)$ and layer $l\in [0, L)$, given the input $\mathbf{X}_{i,l}\in \mathbb{R}^{b\times s\times d}$, where sequence length $s=1$, the router and MoE computation can be formulated as
\begin{gather*}
    \mathbf{A}^{r}_{i,l} = \texttt{Linear}(X_{i,l}),\\
\mathbf{J}_{i,l}, \mathbf{W}_{i,l} = \Phi(\mathbf{A}^{r}_{i,l}), \\
\mathbf{Y}_{i,l} = \texttt{MoE}(\mathbf{X}_{i,l}, \mathbf{W}_{i,l}, E_{l}[\mathbf{J_{i,l}}]),
\end{gather*}
where $\mathbf E_l=\{E_{l,j}\}_{j=0}^{e-1}$ denotes the expert collection of layer $l$. $\mathbf{A}^{r}_{i,l}\in\mathbb{R}^{b\times s\times e}$ denotes the raw \textit{expert activation state}, and $\Phi(\cdot)$ typically consists of softmax followed by a Top-$k$ operation. $\mathbf{J}_{i,l}\in [e]^{b\times s\times k}$, $\mathbf{W}_{i,l}\in \mathbb{R}^{b\times s\times k}$ denote selected expert indices and their routing weights. The output $\mathbf{Y}_{i,l}\in \mathbb{R}^{b\times s\times d}$ is computed by dispatching each token to its selected experts and aggregating their weighted outputs. As shown in Table~\ref{tab:background_moe}, sparse scaling makes MoE layers dominate model parameters while activating only a small fraction of experts per inference step. 

The dynamic sparsity creates a structural advantage for offloading: \textit{only the activated experts need to reside in device memory at each inference step. If they can be loaded in time for computation, offloading can in principle approach FullLoad performance with a much smaller memory footprint}.

\begin{table}[t]
\centering
\caption{Configurations of evaluated MoE models.}
\label{tab:background_moe}
\small

\renewcommand{\arraystretch}{1.2}

\begin{tabular*}{\columnwidth}
{@{\extracolsep{\fill}}lcc@{}}
\toprule
\multirow{2}{*}{\textbf{LLM Model}} 
& \textbf{Parameters} 
& \textbf{Experts per Layer} \\
& \textbf{(MoE / Total)} 
& \textbf{(Active / Total)} \\
\midrule
Qwen3-30B-A3B-FP8     & 29B / 31B   & 8 / 128 \\
Qwen3.6-35B-A3B-BF16  & 32B / 35B     & 8 / 256 \\
GPT-OSS-120B-MXFP4    & 115B / 117B & 4 / 128 \\
DeepSeek-V4-Flash     & 277B / 284B      & 6 / 256 \\
\bottomrule
\end{tabular*}
\end{table}

\subsection{Expert Activation Prediction}
Expert activation is late-bound, making early prediction essential for downstream scheduling. Existing approaches mainly fall into three categories. 

\textit{Similarity-based cross-layer prediction}. ProMoE~\cite{promoe} predicts deeper-layer activations from shallow-layer hidden states with MLPs. Pre-gated~\cite{pregated} advances routing decisions through fine-tuning. MoE-APEX~\cite{moeapex} directly feeds shallow-layer hidden states into future routers. These methods rely on residual-induced cross-layer similarity, causing accuracy to degrade rapidly with prediction distance and limiting lookahead to only one to three layers. For example, on Qwen3-30B-FP8 with an RTX 4090 over PCIe 4.0, one layer’s computation overlaps the transfer of only 1.39 experts on average, leaving little room for prefetching. \textit{Trace-based cross-layer prediction}. MoE-Infinity~\cite{moeinfinity} and Taming-MoE~\cite{tamingmoe} match shallow-layer activation states against request-level and iteration-level histories, respectively. As activation patterns evolve, matching must be updated layer by layer, again limiting lookahead to at most three layers. \textit{First-order cross-step prediction}. ST-MoE~\cite{stmoe} and Patterns-MoE~\cite{patternsmoe} explore cross-step prediction based on per-layer conditional activation statistics between adjacent tokens. This memoryless, first-order Markov model is too coarse for precise scheduling: ST-MoE uses it only to assist cross-layer prediction, while Patterns-MoE confines it to lightweight cache guidance. 

Overall, existing methods cannot simultaneously achieve high prediction accuracy and sufficient lookahead to effectively overlap computation with memory transfers, leaving downstream scheduling with a narrow and unreliable optimization window.

\begin{figure}[t]
  \centering
  \includegraphics[width=\linewidth]{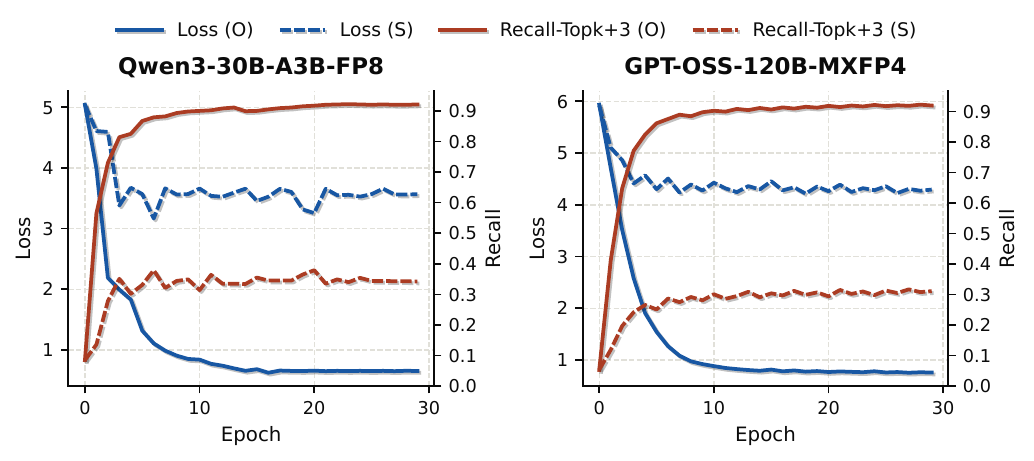}
  \caption{Sequence Dependence in Expert Activation.}
  \Description{...}
  \label{fig:moti_seq}
\end{figure}

\subsection{Sequence Modeling}
Sequence modeling captures dependencies in ordered data to represent or predict states from historical context~\cite{seq2seq}. Given a sequence $x_{1:T}=(x_1,x_2,\dots,x_T)$, it learns a mapping $g:(x_1,\dots,x_t)\mapsto y_t$. MoE expert activations naturally form sequences across generation steps and thus fit this framework. Representative models include Transformers~\cite{transformer}, Recurrent Neural Networks (RNNs)~\cite{rnn}, Long Short-Term Memory (LSTMs)~\cite{lstm}, and recent State Space Models such as Mamba~\cite{mamba2}. 

Transformers offer strong long-range modeling but incur growing Key-Value (KV) cache memory overhead. We therefore use a Transformer only to verify sequence-level dependencies in expert activations and provide an accuracy upper bound. In contrast, RNNs, LSTMs, and Mamba compress history into a fixed-size state:  
\begin{gather*}
    H_{t} = f(H_{t-1}, x_{t}),\quad y_{t} = g(H_{t}, x_{t}),
\end{gather*}
where $f(\cdot)$ and $g(\cdot)$ are the state-update and output functions. The fixed-size state $H_{t}$ enables lightweight prediction with low memory overhead. However, RNNs and LSTMs struggle with long-range dependencies and require sequential training. Mamba enhances long-range modeling through selective, content-aware state updates, while its state-space formulation supports efficient parallel training.

Therefore, Mamba combines a \textit{fixed-size historical state}, \textit{strong long-range modeling capability}, and \textit{efficient parallel training}. We adopt Mamba as the backbone of our sequence-to-sequence (seq2seq) expert activation predictor.

\subsection{Graph-based Execution Mechanisms}
Graph-based execution reduces host-side dispatch overhead by capturing device operations into a graph and replaying them with a single dispatch~\cite{graphexe}. Mechanisms such as NVIDIA CUDA Graphs~\cite{nvidiagraph}, AMD HIP Graphs~\cite{amdgraph}, Ascend ACL Graphs~\cite{huaweigraph}, and Intel SYCL Graphs~\cite{intelgraph} are particularly effective for small-batch LLM decoding, where short kernels make dispatch overhead significant. Edge inference typically operates at small batch sizes and thus benefits substantially. For MoE offloading, as on-demand loading stalls recede, reducing dispatch overhead becomes critical to approaching full-load performance.

Efficient graph replay favors fixed execution structures and memory addresses while minimizing host intervention and synchronization. These requirements conflict with MoE offloading. \textit{First, dynamic expert placement hinders the use of fused MoE kernels, a key enabler of graph-based MoE execution}. Although kernels such as Triton Fused MoE~\cite{tritonfusedmoe} and DeepGEMM~\cite{deepgemm} encapsulate dynamic activation within fixed execution structures, offloading dynamically maps experts to cache slots, making their addresses unstable. \textit{Second, prefetching and caching traditionally rely on CPU-GPU synchronization, which breaks graph execution and re-exposes dispatch overhead}. FreeToken~\cite{freetoken}, a recent concurrent work, moves LRU cache management to the GPU to adapt offloading to CUDA Graph execution. However, it does not support prefetching, a basic offloading operation, nor does it derive general principles for graph-compatible offloading. 

The offloading runtime must therefore be redesigned to coordinate prediction, prefetching, and caching with LLM inference under graph execution constraints. 


\begin{table}[b]
\caption{Prefetch Scheduling Example}
\label{tab:moti_prefetch}
\renewcommand\arraystretch{1.05}
\setlength\tabcolsep{5.3pt}

\begin{tabular}{
!{\vrule width 1.2pt}
c|c|cc|ccc
!{\vrule width 1.2pt}
}
\specialrule{1.2pt}{0pt}{0pt}

\textbf{Candidates}
& A
& \multicolumn{1}{c|}{B}
& C
& \multicolumn{1}{c|}{D}
& \multicolumn{1}{c|}{E}
& F
\\ \hline

\textbf{Deadline}
& 1
& \multicolumn{1}{c|}{2}
& 2
& \multicolumn{1}{c|}{3}
& \multicolumn{1}{c|}{3}
& 3
\\ \hline

\textbf{Confidence}
& 0.55
& \multicolumn{1}{c|}{0.90}
& 0.40
& \multicolumn{1}{c|}{0.75}
& \multicolumn{1}{c|}{0.70}
& 0.30
\\ \hline

\textbf{Capacity}
& 1
& \multicolumn{2}{c|}{3}
& \multicolumn{3}{c!{\vrule width 1.3pt}}{4}
\\
\specialrule{1.2pt}{0pt}{0pt}

\rowcolor[HTML]{EFEFEF}
{\color[HTML]{333333}\textit{\textbf{Optimal}}}
& {\color[HTML]{333333}\textit{\textbf{A}}}
& \multicolumn{1}{c|}{
    \cellcolor[HTML]{EFEFEF}
    {\color[HTML]{333333}\textit{\textbf{B}}}
  }
& {\color[HTML]{333333}\textit{\textbf{D}}}
& \multicolumn{3}{c!{\vrule width 1.2pt}}{
    \cellcolor[HTML]{EFEFEF}
    {\color[HTML]{333333}\textit{\textbf{E}}}
  }
\\ \hline

\textbf{MoE-Infinity}
& B
& \multicolumn{1}{c|}{D}
& E
& \multicolumn{3}{c!{\vrule width 1.2pt}}{F}
\\ \hline

\textbf{ProMoE}
& A
& \multicolumn{1}{c|}{B}
& C
& \multicolumn{3}{c!{\vrule width 1.2pt}}{D}
\\
\specialrule{1.2pt}{0pt}{0pt}

\end{tabular}
\end{table}

\section{Motivation and Key Idea}
\subsection{Expert Activation Sequence Modeling}
\label{sec:Expert Activation Sequence Modeling}
\textbf{Expert activation exhibits strong sequence-level dependence that can be effectively exploited through sequence modeling}. We encode expert activations across the full layer stack at each generation step into a vector and arrange these vectors in generation order to form an activation sequence. We train decoder-only Transformers for next-step prediction on the original (O) and shuffled (S) sequences, respectively. Fig.~\ref{fig:moti_seq} reveals two observations. First, training fails to converge after shuffling, indicating that predictability stems from sequence-level dependence in expert activations rather than merely model capacity, demonstrating the sufficiency of sequence modeling. Second, top-11 recall on Qwen3-30B-A3B ($e=128,k=8$) and top-7 recall on GPT-OSS-120B ($e=128,k=4$) both exceed 90\%, substantially outperforming existing cross-step prediction methods (Section~\ref{sec:Prediction Accuracy}), demonstrating the necessity of explicitly modeling such sequential dependence.

\subsection{Joint Structure in Prefetch Scheduling}
\label{Joint Structure in Prefetch Scheduling}
\textbf{Bandwidth scarcity turns prefetch scheduling into a constrained task-selection problem}. A slow interconnect can transfer only a subset of prefetch candidates, which differ along two dimensions: prediction confidence, the likelihood that an expert will be activated, and deadline, the time until its owning layer executes. Let $\texttt{ddl}_t,\texttt{pos}\in[0,I\times L)$ denote the deadline of task $t$ and the current position, respectively. $\texttt{conf}_t$ denotes its expected hit contribution, and $C(\texttt{ddl})$ the transfer capacity available before $\texttt{ddl}$. Maximizing expert hits can then be formulated as:
\begin{gather*}
    \max_{T}\sum_{t\in T} \texttt{conf}_{t}\ s.t.\ |\{t\in T:\texttt{ddl}_t\leq \texttt{ddl}\}|\leq C(\texttt{ddl}),\forall \texttt{ddl}.
\end{gather*}
MoE-Infinity prioritizes tasks by $\texttt{conf}_t(1-(\texttt{ddl}_t-\texttt{pos})/(L+1))$, with $\texttt{pos}$ advancing dynamically, while ProMoE sorts tasks lexicographically by $(-\texttt{ddl}_t,\texttt{conf}_t)$. As Table~\ref{tab:moti_prefetch} shows, neither yields an optimal selection. Both rank tasks independently, overlooking how their selections jointly consume transfer capacity under deadline constraints. The resulting bandwidth waste can grow as longer overlap windows expand the scheduling space. In contrast, the feasible task sets form a matroid, allowing weighted matroid greedy to optimally solve the formulated problem~\cite{matroids}. We use this formulation to guide online prefetch scheduling and improve bandwidth efficiency.

\begin{figure}[t]
  \centering
  \includegraphics[width=\linewidth]{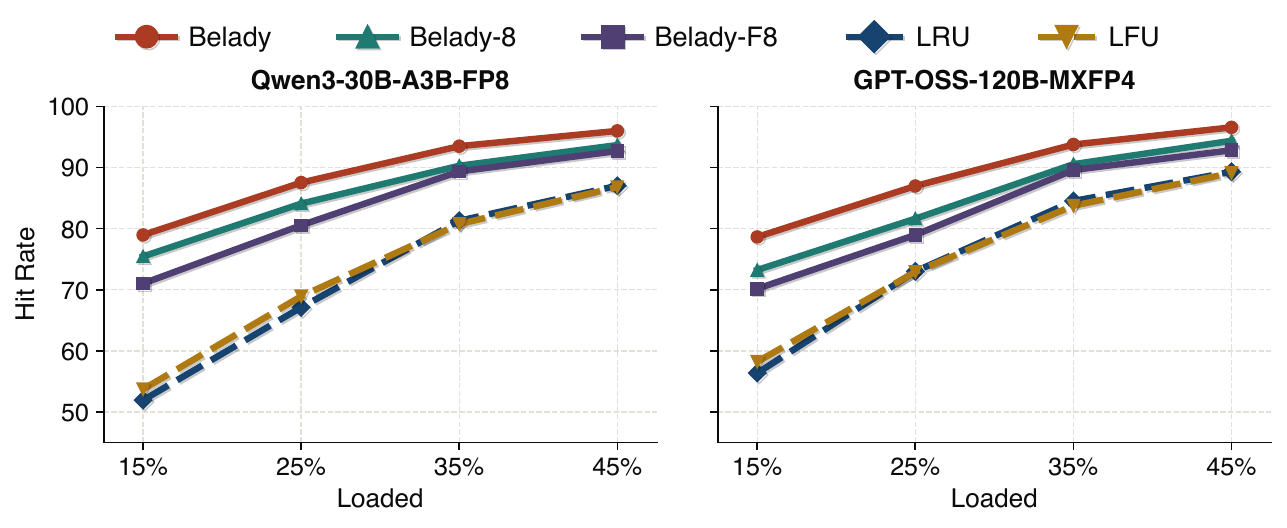}
  \caption{Benefits of Future-Aware Eviction.}
  \Description{...}
  \label{fig:moti_cache}
\end{figure}

\subsection{Future-Aware Cache Eviction}
\label{sec:Future-Aware Cache Eviction}
\textbf{Future-aware eviction consistently outperforms history-based policies}. Existing approaches typically adapt classical eviction policies such as LRU and LFU. Fig.~\ref{fig:moti_cache} compares them with variants of the optimal Belady’s MIN~\cite{beladymin}, where the x-axis denotes the fraction of experts that can reside in the cache. Belady uses the complete future activation sequence and evicts the expert whose next activation is farthest in the future. Belady-8 limits this oracle knowledge to the next eight steps, while Belady-F8 replaces oracle activations with our eight-step forecasts (Section~\ref{sec:Probabilistic Belady Policy}). We make three observations. First, LRU and LFU perform similarly, suggesting that both capture only part of the expert reuse patterns. Second, future-aware policies consistently outperform history-based ones, highlighting the value of future activation information for eviction. Third, Belady-F8 closely approaches Belady-8, showing that predicted activations provide a reliable basis for cache eviction.

\begin{figure}[t]
  \centering
  \includegraphics[width=\linewidth]{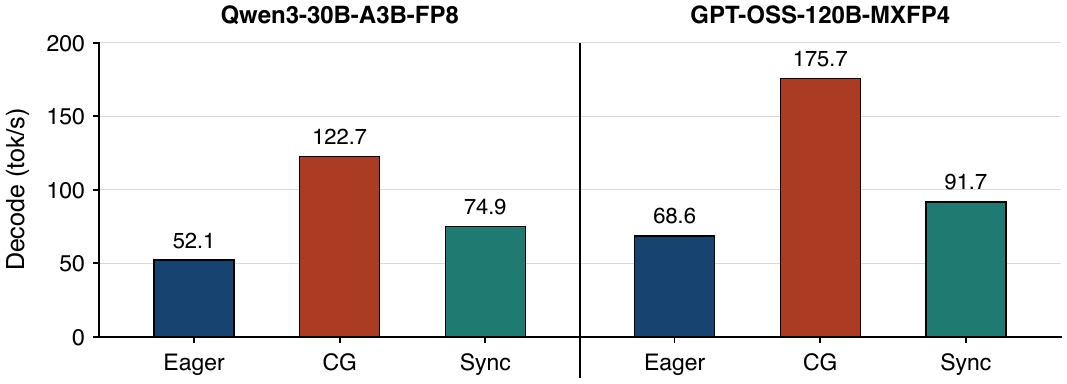}
  \caption{Impact of End-to-End Graph Capture.}
  \Description{...}
  \label{fig:moti_cg}
\end{figure}

\subsection{End-to-End Graph Execution}
\label{sec:End-to-End Graph Execution}
\textbf{As on-demand loading stalls recede, maintaining low dispatch overhead becomes critical to approaching full-load performance}. We place all model weights in device memory to simulate a 100\% expert hit rate and isolate the performance impact of graph-based execution. Fig.~\ref{fig:moti_cg} compares three execution modes. Eager represents existing runtimes that cannot accommodate dynamic expert placement in fused MoE kernels and instead dispatch individual GEMMs according to expert activation. CG combines fused MoE kernels with CUDA Graphs, representing ideal full-load execution. Sync breaks the graph between attention and MoE to reflect CPU–GPU synchronization introduced by prefetching and caching operations in existing runtimes. Both eager execution and synchronization substantially degrade overall inference performance. These results highlight the necessity of end-to-end graph capture for achieving high-performance MoE offloading inference.


\section{Overview}
Fig.~\ref{fig:overview} presents an overview of SeqMoE. During inference, each Transformer block enters the offloading pipeline after attention and routing. The system first lands previously issued prefetch tasks, which may target experts from any layer, and updates cache slot states (Section~\ref{sec:Slot State Transition}). It then loads any missing experts required by the current layer from host to device memory. At a predictor’s trigger layer $l'$, the predictor forecasts expert activation probabilities over multiple steps for its assigned layers $l$ to $l'$ (Section~\ref{sec:Seq2Seq Expert Activation Prediction}). The first-step probabilities generate prefetch tasks (Section~\ref{sec:Prefetch Task Generation}), which the prefetcher schedules as a job sequencing with deadlines problem using matroid greedy (Section~\ref{sec:Prefetch Task Selection}). Subsequent-step probabilities update the probabilistic Belady policy for future-aware eviction (Section~\ref{sec:Probabilistic Belady Policy}). Finally, the runtime’s compute-transparent expert placement allows execution to leave the offloading pipeline and resume fused MoE computation (Section~\ref{sec:Compute-Transparent Expert Placement}). Beyond this pipeline, we establish four runtime orchestration disciplines that govern how these memory-management components coordinate to enable graph-based execution (Section~\ref{sec:Offloading Orchestration Discipline}).

\section{Seq2Seq Expert Activation Prediction}
\label{sec:Seq2Seq Expert Activation Prediction}
\subsection{Expert Activation Collection}
We collect expert activation traces across diverse workloads, including mathematics (GSM8K~\cite{gsm8k}, MATH~\cite{math}), code (CodeForces~\cite{codeforces}), text comprehension (OpenOrca~\cite{openorca}), and multi-turn conversations (ShareGPT~\cite{sharegpt}). To improve collection efficiency, we adopt a two-pass process. The first pass collects token ID traces through continuous batching. The second packs multiple token ID traces into a single prefill pass to dump expert activations, fully exploiting the inference engine’s parallelism. We denote the expert activations spanning multiple steps and layers as
\begin{gather*}
    \mathbf{A}^{r}_{i:i',l:l'} = \texttt{Stack}_{k=l}^{l'-1}(\texttt{Concat}_{j=i}^{i'-1} \mathbf{A}^{r}_{j,k})\in \mathbb{R}^{b\times s\times n\times e},
\end{gather*}
where $s=i'-i$ is the sequence length and $n=l'-l$ is the number of layers. Each collected expert activation trace is represented as $\mathbf{A}^{r}_{0:I,0:L}$ $\in$ $\mathbb{R}^{b\times I \times L \times e}$. 

\begin{figure*}[t]
  \centering
  \includegraphics[width=\textwidth]{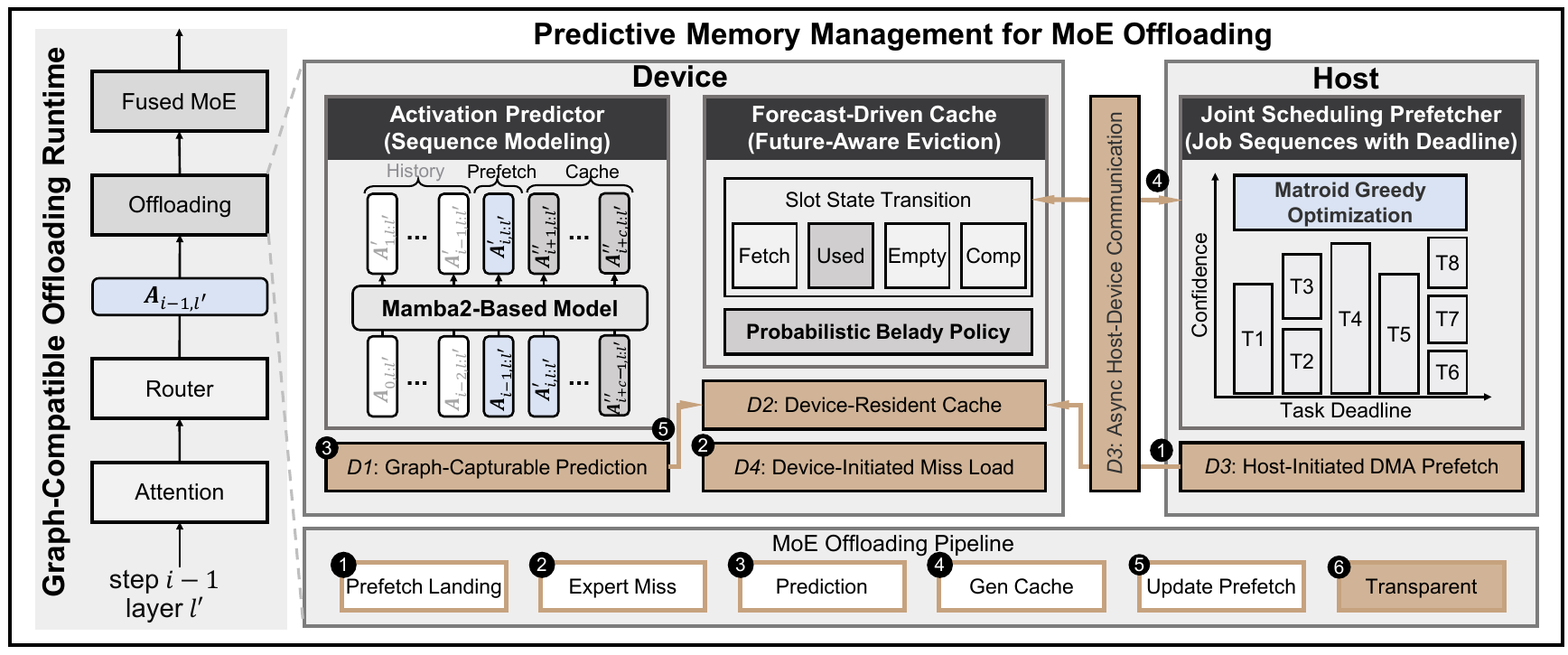}
  \caption{SeqMoE System Overview.}
  \Description{...}
  \label{fig:overview}
\end{figure*}

\subsection{Model Architecture}
We next introduce the preprocessing and model architecture of the predictor. The preprocessing is defined as
\begin{gather*}
    \mathbf{A}_{i:i',l:l'} = \Psi(\mathbf{A}^{r}_{i:i',l:l'}) \in \mathbb{R}^{b\times s \times (n \times e)},
\end{gather*}
where $\Psi(\cdot)$ consists of two steps. First, activation-masked normalization sets inactive experts to $\texttt{-inf}$ and applies softmax along the expert dimension. This preserves the relative preference among activated experts as prediction confidence for prefetch scheduling, while suppressing noise from irrelevant experts to improve recursive prediction and provide more reliable signals for cache eviction. Second, reshaping merges the layer and expert dimensions into a single feature dimension, treating activations across multiple layers as a unified representation at each generation step. The number of included layers introduces a trade-off between prediction accuracy and prefetch overlap: including more layers improves accuracy but shortens the overlap window.

Based on the preprocessed input, the predictor can be formulated as
\begin{gather*}
    \mathbf{Z}_1 = \texttt{Linear}(\mathbf{A}_{i:i',l:l'})\in \mathbb{R}^{b \times s\times m}, \\
\mathbf{Z}_2,\mathbf{H}_{i'} = \texttt{Mamba2}(\mathbf{Z}_1, \mathbf{H}_{i}), \\
\mathbf{A}'_{i+1:i'+1, l:l'}=\texttt{Linear}(\mathbf{Z_{2}}) \in \mathbb{R}^{b \times s \times (n\times e)},
\end{gather*}
where $m$ denotes hidden dimension of Mamba2. Its learned recurrent state accumulates information from the $i'-i$ input activation states, updating from $\mathbf{H}_{i}$ to $\mathbf{H}_{i'}$. Each output in $\mathbf{A}'_{i+1:i'+1,l:l'}$ predicts the activation probabilities for the step immediately following its corresponding input.

\subsection{Prediction Execution}
The predictor supports four execution modes. Consider a predictor responsible for layers $[l,l')$: 

\textbf{Training}. The predictor takes $\mathbf{A}_{0:I-1,l:l'}$ $\in$ $\mathbb{R}^{b\times (I-1) \times (n \times e)}$ as input and produces $\mathbf{A}'_{1:I,l:l'}$, supervised by the corresponding ground truth $\mathbf{A}_{1:I,l:l'}$. No recurrent state needs to be retained across training sequences. 

\textbf{Prefill}. For a prompt length $p$, the predictor consumes $\mathbf{A}_{0:p,l:l'}$ $\in$ $\mathbb{R}^{b\times p \times (n \times e)}$ and outputs $\mathbf{A}'_{1:p+1,l:l'}$, while retaining the recurrent state $\mathbf{H}_{p}$. Predictor prefill proceeds along with LLM prefill, and only the final prediction $\mathbf{A}'_{p,l:l'}$ is used for downstream operations. 

\textbf{One-step prediction}. Suppose LLM inference reaches step $i-1$ at layer $l'$. Given $\mathbf{H}_{i-1}$, the predictor takes $\mathbf{A}_{i-1,l:l'}$ $\in$ $\mathbb{R}^{b\times 1 \times (n \times e)}$ and produces $\mathbf{A}'_{i,l:l'}$, updating the state to $\mathbf{H}_i$. This mode runs at every generation step to issue prefetch tasks for the next step. 

\textbf{Recursive prediction}. Starting from the one-step prediction above, the predictor repeatedly feeds its latest output back as input. The first recursive input is $\Psi(\mathbf{A}'_{i,l:l'})$. After $c$ iterations, it produces $\mathbf{A}''_{i+1:i+c+1,l:l'}$ $\in$ $\mathbb{R}^{b\times c\times (n\times e)}$, which forecasts expert activations over the next $c$ steps. The recurrent state is then restored to $\mathbf{H}_i$ so that recursive rollout does not affect the next one-step prediction. Recursive prediction also runs at every generation step to update the cache policy.

\subsection{Training}
We combine a ranking loss~\cite{rankingloss} with KL divergence loss~\cite{klloss}. Since inference selects predicted experts by Top-$k$, the ranking loss directly optimizes this decision: it enforces ground-truth Top-$k$ experts to score higher than hard negatives by a margin, improving the ordering around the Top-$k$ decision boundary. We further apply KL divergence between the prediction $\mathbf{A}'_{1:I,l:l'}$ and ground truth $\mathbf{A}_{1:I,l:l'}$. Since the ground-truth distribution follows the same space as the predictor input $\mathbf{A}_{0:I-1,l:l'}$, this objective aligns the input and output distributions, facilitating recursive prediction.

To further improve recursive prediction, we first train the predictor with teacher forcing, where ground-truth activations are always used as inputs for subsequent predictions. After convergence, we adopt Scheduled Sampling~\cite{scheduledsampling} by probabilistically replacing a subset of $\mathbf{A}_{0:I-1,l:l'}$ with the corresponding predictions from $\mathbf{A}'_{1:I,l:l'}$ and continue training. This gradually exposes the predictor to its own outputs during training, improving its robustness to error accumulation in recursive inference. 

\section{Joint Prefetch Scheduling}
\subsection{Prefetch Task Generation}
\label{sec:Prefetch Task Generation}
Suppose inference proceeds to step $i-1$, layer $l'$, where the predictor performs one-step prediction and outputs the expert activation probabilities $\mathbf{A}'_{i,l:l'}$ $\in$ $\mathbb{R}^{b\times 1 \times n \times e}$. Based on this prediction, the runtime generates $k\leq k' < e$ prefetch tasks for experts with high activation probabilities. Consider a generated task $t$ corresponding to expert $E_{p,q}$, where $l \leq p < l'$ denotes its layer and $0\leq q < e$ denotes its expert index. Its prediction confidence, which quantifies its expected contribution to expert hits, and deadline, which indicates when the expert is needed for computation, are defined as
\begin{gather*}
    \texttt{conf}_{t} = \texttt{Mean}(\mathbf{A}'_{i,l:l'}[:, 0, p, q]),\\
\texttt{ddl}_{t} = (i-1)\cdot L + p
\end{gather*}
Accordingly, task $t$ has an overlap window of $L-(l'- 1 - p)$, during which the expert transfer can overlap with ongoing inference. Within this window, the task remains a candidate for scheduling and, once selected, performs the actual expert transfer from the next-level memory to device memory.

\begin{algorithm}[t]
\caption{Joint Prefetch Scheduling}
\label{alg:prefetch}
\renewcommand{\baselinestretch}{1.07}\selectfont
\begin{algorithmic}[1]
\Statex\hspace{-\algorithmicindent} \textbf{Inputs:} 
$Q_\texttt{task}$: current prefetch task queue, 
$\texttt{pos}$: current inference position, 
$B_\texttt{avg}$: average transfer budget per layer, 
$B_\texttt{cur}$: remaining transfer budget of current layer.

\Statex\hspace{-\algorithmicindent} \textbf{Output:} 
$\texttt{selected}$: selected task for prefetching.

\State \textbf{if} $\texttt{pos}$ is updated \textbf{then}
\State \hspace*{0.5em} $B_\texttt{avg} \gets B_\texttt{avg} + \alpha \times B_\texttt{cur}$
\State \hspace*{0.5em} $B_\texttt{cur} \gets B_\texttt{avg}$
\State \hspace*{0.5em} $Q_\texttt{task} \gets
\{t \in Q_\texttt{task} \mid \texttt{ddl}_{t} > \texttt{pos}+1\}$

\State Sort $Q_\texttt{task}$ by $\texttt{ddl}_{t}$ in ascending order
\State Initialize an empty min-heap $\texttt{Heap}$

\State \textbf{for} $t$ \textbf{in} $Q_\texttt{task}$ \textbf{do}
\State \hspace*{0.5em} $\texttt{Heap}.\textsc{Push}(t;\ \texttt{conf}_{t}, -\texttt{ddl}_{t})$
\State \hspace*{0.5em} \textbf{if} $|\texttt{Heap}| > C(\texttt{ddl}_{t})$ \textbf{then}
\State \hspace*{1.3em} $\texttt{Heap}.\textsc{Pop}()$

\State $\texttt{selected} \gets
\arg\min_{t \in \texttt{Heap}} \texttt{ddl}_{t}$
\State $\texttt{B}_{cur} \gets \texttt{B}_{cur} - 1$
\State \textbf{return} $\texttt{selected}$

\end{algorithmic}
\end{algorithm}

\subsection{Prefetch Task Selection}
\label{sec:Prefetch Task Selection}
Algorithm~\ref{alg:prefetch} shows how the current prefetch task is selected. The average transfer budget $B_{\texttt{avg}}$ is profiled during warmup. When inference advances to a new layer (Line 1), we update $B_{\texttt{avg}}$ with an Exponential Moving Average (EMA) to adapt to runtime bandwidth variation and reset the current-layer budget $B_{\texttt{cur}}$ accordingly (Lines 2–3). We then remove tasks whose deadlines are no later than the next layer, as they are either already expired or may not complete in time (Line 4). 

The remaining tasks are processed in ascending deadline order to construct the maximum-confidence feasible set under the deadline constraints (Lines 5–10). Specifically, for a task with deadline $\texttt{ddl}$, the available transfer capacity before its deadline is
\begin{gather*}
    C(\texttt{ddl}) = \max(1,B_{\texttt{cur}}) + B_{\texttt{avg}} \times (\texttt{ddl} - \texttt{pos}-1),
\end{gather*}
which accounts for the remaining budget of the current layer and the transfer capacity of subsequent layers. Following matroid greedy, tasks are inserted into a min-heap keyed by confidence, and the lowest-confidence task is removed whenever the retained set exceeds $C(\texttt{ddl})$, ensuring feasibility while prioritizing tasks with higher expected contributions to expert hits. This preserves the highest-confidence feasible task set under all deadline constraints. Finally, Earliest Deadline First (EDF) selects the task with the earliest deadline for transfer (Line 11). The prefetcher waits for the transfer to complete and then reruns the algorithm to select the next prefetch task.

\section{Forecast-driven Caching}
\subsection{Probabilistic Belady Policy}
\label{sec:Probabilistic Belady Policy}
Suppose inference reaches step $i-1$ at layer $l'$. After completing the single-step prediction, the predictor recursively generates $\mathbf{A}''_{i+1:i+c+1,l:l'}$, which estimates expert activation probabilities over the next $c$ steps. The optimal Belady's MIN retains objects that will be reused sooner and evicts those whose next reuse lies farthest in the future. Inspired by this principle, we propose a probabilistic Belady policy that extends future reuse to predicted activation probabilities. For expert $E_{p,q}$, where $l\leq p<l'$ denotes its layer and $0\leq q<e$ its expert index, we define its retention priority as
\begin{gather*}
    \texttt{CacheScore}_{p,q} = \frac{\sum_{j=0}^{c-1}(\lambda^{j}\cdot \texttt{Mean}(\mathbf{A}''[:, i, p, q]))}{\sum_{j=0}^{c-1}\lambda^{j}},
\end{gather*}
where $0< \lambda < 1$. This score prioritizes experts with higher activation probabilities in the nearer future, approximating Belady’s preference for earlier reuse.

\subsection{Slot State Transition}
\label{sec:Slot State Transition}
Each cache slot has one of four states. $\texttt{Empty}$ denotes a free slot that can be directly allocated without eviction. $\texttt{Used}$ holds a resident expert that is eligible for eviction. $\texttt{Fetch}$ denotes a slot reserved for an ongoing prefetch and is protected from eviction. $\texttt{Compute}$ holds an expert awaiting computation and is also protected from eviction.

Four events in the offloading pipeline trigger slot-state transitions. First, when creating a prefetch task, an $\texttt{Empty}$ or $\texttt{Used}$ slot is reserved and transitioned to $\texttt{Fetch}$. Second, when the prefetch lands, the slot transitions from $\texttt{Fetch}$ to $\texttt{Compute}$ if the transfer completes successfully. Otherwise, it falls back to $\texttt{Empty}$. Third, an on-demand load directly transitions an allocated $\texttt{Empty}$ or $\texttt{Used}$ slot to $\texttt{Compute}$. Finally, immediately before executing each MoE layer, all $\texttt{Compute}$ slots belonging to that layer are transitioned to $\texttt{Used}$, including both activated experts and prefetched-but-unactivated experts, making them eligible for subsequent eviction.

Following these transitions, slot allocation always prioritizes $\texttt{Empty}$ slots. If none is available, we evict the expert with the lowest $\texttt{CacheScore}$ among $\texttt{Used}$ slots.

\section{Graph-Compatible Offloading Runtime}
\subsection{Compute-Transparent Expert Placement}
\label{sec:Compute-Transparent Expert Placement}
Fused MoE kernels are a key enabler of graph-based MoE execution, as they encapsulate dynamic expert activation within a fixed execution structure. They typically require each layer’s expert weights to follow a fixed layout:
\begin{gather*}
    [\texttt{num\_experts},\ \texttt{dim}_{0},\ \texttt{dim}_{1}],
\end{gather*}
where experts are stored contiguously at a fixed base address and selectively computed according to $\texttt{expert\_id}$. Offloading breaks this structure by dynamically placing experts at arbitrary locations.

Our key idea is to share cache slots across all MoE layers, whose expert-weight pointers reference the same fixed base address with the layout:
\begin{gather*}
    [\texttt{num\_slots},\ \texttt{dim}_{0},\ \texttt{dim}_{1}].
\end{gather*}
This design offers four benefits. First, it restores the fixed base address required by fused MoE kernels. Once all activated experts are cached, a lookup table maps $\texttt{expert\_id}$ to physical $\texttt{slot\_id}$, preserving computation semantics. Second, although $\texttt{num\_slots}$ can far exceed $\texttt{num\_experts}$, sparse activation keeps kernel latency unchanged in our measurements, preserving computation performance. Third, global sharing improves the flexibility of caching and prefetch scheduling. Finally, it accommodates the many experts activated during prefill without the per-layer capacity limits of statically partitioned slots.

This design fully decouples the offloading pipeline from model computation, restoring fused-kernel efficiency and a fixed execution structure amenable to graph capture.

\subsection{Offloading Orchestration Discipline}
\label{sec:Offloading Orchestration Discipline}
We further eliminate host-device synchronization in the offloading pipeline to incorporate the pipeline itself into graph-based execution. As shown in Fig.~\ref{fig:overview}, we introduce four orchestration disciplines for its core components: the predictor, cache, prefetcher, and on-demand loading.
\begin{itemize}
    \item \textbf{Graph-capturable prediction}. The predictor must run entirely on the device, and its prediction process can be directly captured into the graph.
    \item \textbf{Device-resident cache}. Cache management must run entirely on the device, avoiding frequent host-device synchronization for expert access information and cache decisions. 
    \item \textbf{Device-initiated miss loading}. On a cache miss, the device must initiate the required data transfer and track its progress, avoiding host-device synchronization to determine transfer completion.
    \item \textbf{Asynchronous host-device communication and host-initiated DMA}. The prefetcher runs on the CPU for scheduling flexibility, while both control communication and host-initiated data transfers must remain fully asynchronous.
\end{itemize}
Together, these disciplines coordinate prediction, cache management, prefetch scheduling, and on-demand loading without introducing host-device synchronization into the execution path. This coordination enables end-to-end graph capture of the offloading pipeline while preserving flexible host-side scheduling and asynchronous data movement.

\subsection{Implementation}
Following these disciplines, our implementation uses two threads. The main thread issues only CUDA operations for model inference, prediction, cache management, and on-demand loading, while a CPU worker thread runs prefetcher. CUDA Graph captures the main thread’s CUDA operations, with prefetcher activity fully overlapped.

For prediction, our seq2seq predictor has no dynamic control flow, allowing direct capture of both one-step and recursive prediction. For cache management, all metadata, including lookup and state tables, reside on the GPU, with operations such as lookup and eviction implemented as CUDA kernels.  For on-demand loading, we allocate the host expert pool with $\texttt{cudaHostAllocMapped}$. A custom kernel on the main thread’s CUDA stream copies missing experts from host to GPU memory, avoiding host synchronization while remaining graph-capturable.

For prefetching, we similarly use $\texttt{cudaHostAllocMapped}$ to allocate pinned, mapped host memory as a shared control region between CPU and GPU, supporting three forms of asynchronous communication. First, the GPU updates the inference position $\texttt{pos}$ at each layer, which the CPU prefetcher reads after completing each prefetch task. Second, when a cache miss triggers on-demand loading, the GPU sets a shared flag to pause new prefetch submissions, reducing bandwidth contention. Third, monotonically increasing tags coordinate task dispatch and prefetch landing. The GPU assigns each new task a slot and binds its tag to that slot, and the CPU processes only unseen tags. The GPU updates the cache state only when the returned tag matches the slot’s expected tag. Prefetched weights bypass the control region: the CPU prefetcher initiates host-to-device DMA using $\texttt{cudaMemcpyAsync}$ on a dedicated CUDA copy stream. Consequently, prefetch configuration, task generation, data transfer, and completion notification proceed asynchronously with the main computation stream, and GPU execution never waits for the CPU prefetcher.

We implement SeqMoE in Hugging Face Transformers~\cite{huggingfacetransformers} using hooks. A single offloading function, inserted after routing and before MoE computation, incorporates these operations without changing external interfaces.

\subsection{Prefill Offloading}
Prefill rarely achieves full-load performance under offloading because many tokens activate a broad set of experts, substantially weakening MoE’s structural advantage from sparsity. Chunked prefill, a common optimization in cloud serving~\cite{chunkedprefill}, is poorly suited to offloading because it incurs repeated expert transfers across chunks.

With many tokens routed simultaneously, expert activation increasingly reflects statistical popularity. We therefore profile expert activation frequencies offline over the dataset. Before prefill and after each layer, we prioritize prefetch tasks for high-frequency activated experts in subsequent layers. We keep $\texttt{num\_slots}-e$ slots in the $\texttt{Fetch}$ state, reserving $e$ slots for evicting and loading missed experts. We omit graph capture during prefill because of its variable sequence lengths and substantial computation, consistent with common inference practice~\cite{vllm}.

\begin{figure*}[t]
  \centering
  \includegraphics[width=\textwidth]{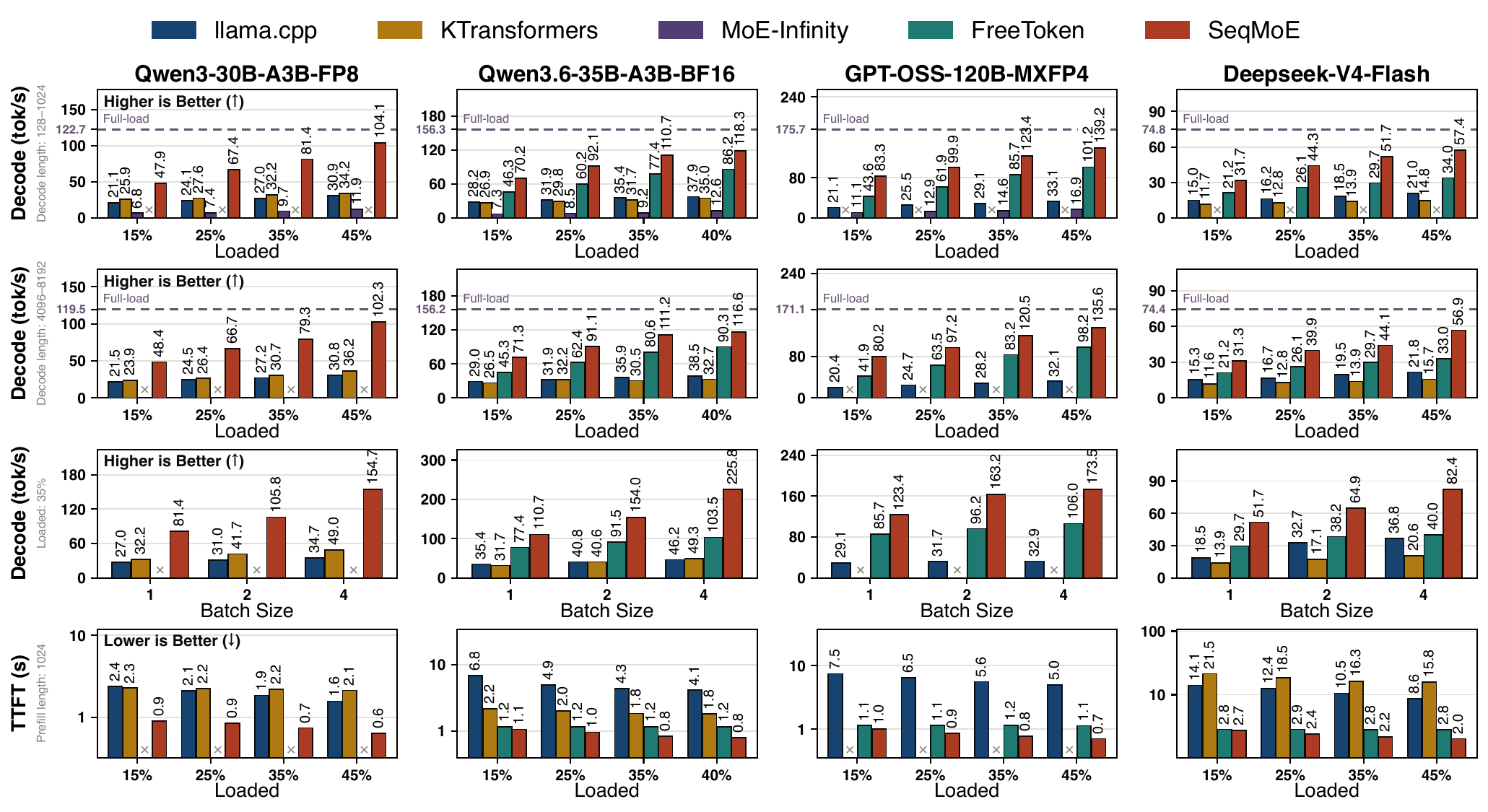}
  \caption{End-to-End Inference Performance Comparison.}
  \Description{...}
  \label{fig:ex_performance}
\end{figure*}

\section{Experimental Evaluation}
\subsection{Experimental Methodology}
\noindent \textbf{Hardware}. We evaluate on three GPU-based platforms. The NVIDIA RTX 4090 platform has 24 GB GPU memory, an Intel Xeon Gold 6430 CPU, and 120 GB host memory, connected via PCIe 4.0. The NVIDIA RTX 5090 platform has 32 GB GPU memory, an Intel Xeon Platinum 8470Q CPU, and 120 GB host memory, connected via PCIe 5.0. The NVIDIA RTX PRO 6000 Blackwell platform has 96 GB GPU memory, an Intel Xeon Platinum 8470Q CPU, and 256 GB host memory, connected via PCIe 5.0.

\noindent \textbf{Models}. We evaluate four advanced, widely used MoE-based LLM models with diverse architectures and precisions: Qwen3-30B-A3B-FP8 (QW3)~\cite{yang2025qwen3}, GPT-OSS-120B-MXFP4 (GPT)~\cite{openai2025gptoss}, Qwen3.6-35B-A3B-BF16 (QW36)~\cite{qwen2026qwen36}, and DeepSeek-V4-Flash (DSV4)~\cite{deepseek2026v4}. Table~\ref{tab:background_moe} summarizes their model weight configurations and expert activation characteristics.

\noindent \textbf{Datasets}. For each model, we sample 5K, 5K, 5K, 30K, and 5K examples from MATH~\cite{math}, GSM8K~\cite{gsm8k}, CodeForces~\cite{codeforces}, OpenOrca~\cite{openorca}, and ShareGPT~\cite{sharegpt}, respectively, to construct a pool of 50K traces covering mathematics, coding, general text comprehension, and multi-turn dialogue, with sequence lengths of 0.1K–10K tokens.

\noindent \textbf{Configuration}. We use 90\% of each subset, totaling 45K traces, for predictor training and the remaining 5K for predictor testing. We further select 600 test traces for inference evaluation. We evaluate QW3 on the NVIDIA RTX 4090, QW36 on the NVIDIA RTX 5090, and both GPT and DSV4 on the NVIDIA RTX PRO 6000 Blackwell. During inference, we generate $k'=k+3$ prefetch tasks per prediction and use $c=8$ recursive prediction steps with $\lambda=0.5$ for the cache policy. We deploy two predictors, triggered at $l'=L/2$ and $l'=L$, respectively.

\noindent \textbf{Baselines}. We compare SeqMoE’s inference performance against four frameworks. Llama.cpp~\cite{llamacpp} supports concurrent CPU–GPU computation in C/C++. MoE-Infinity~\cite{moeinfinity} uses request-level expert activations for GPU prefetching and caching. KTransformers combines CPU–GPU hybrid inference with optimized CPU expert kernels. FreeToken~\cite{freetoken}, a concurrent work, supports graph-based offloading without prefetching and uses LRU without explicit expert-hit optimization. FullLoad keeps all weights on GPUs, using pipeline parallelism (PP) when multiple GPUs are needed for fairness with offloading. It runs on vLLM~\cite{vllm} 0.28.0 for DSV4 and 0.26.0 for other models. For expert hit rates, we compare the caching policies of these frameworks and ablate our prefetch scheduling. For prediction accuracy, we compare against Patterns-MoE~\cite{patternsmoe}, the first work to explore cross-step expert prediction using conditional probabilities.

\noindent \textbf{Metrics}. Tokens per second (tokens/s) measures decoding throughput, while time to first token (TTFT, s) measures prefill latency. Expert hit rate (\%) is the fraction of required experts already resident in device memory before on-demand loading for MoE computation, indicating memory management efficiency and the potential to approach FullLoad performance. Recall (\%) measures prediction accuracy as the fraction of actually activated experts correctly predicted.

\subsection{End-to-End Performance}
Fig.~\ref{fig:ex_performance} compares end-to-end performance. Crosses indicate unsupported configurations. We evaluate MoE-Infinity only in the first experiment group due to its low inference speed. The x-axis shows the number of available device cache slots as a fraction of all experts, ranging from 15\% to 40\% for QW36 due to RTX 5090 memory limits, and from 15\% to 45\% for the other three models.

Continuous serving at the edge often focuses on inference with a batch size of one~\cite{moeapex,freetoken,powerinfer2}. The first row reports performance on shorter sequences (128–1,024 tokens). At 45\% cache capacity (40\% for QW36), SeqMoE achieves 84.82\%, 75.69\%, 79.20\%, and 76.65\% of FullLoad performance on QW3, QW36, GPT, and DSV4, respectively. FreeToken reaches only 55.15\%, 57.60\%, and 45.45\% on QW36, GPT, and DSV4, while the remaining baselines achieve at most 28.03\%. Moreover, SeqMoE’s high expert hit rate enables it to match FreeToken at 45\% or 40\% capacity using only 25\%. The second row reports results on longer sequences (4,096–8,192 tokens). Increased computation and hit-rate fluctuations cause minor performance changes, but the overall ranking persists as SeqMoE maintains prediction accuracy over long sequences.

The third row evaluates small-batch inference with varying numbers of concurrent requests. SeqMoE achieves the largest performance gains as concurrency increases, since competition among sequences for limited cache capacity reduces expert reuse and makes timely prefetching increasingly important. The fourth row compares prefill latency across cache capacities. FreeToken uses ping-pong buffers that alternate between layers, resulting in largely stable latency as cache capacity increases. In contrast, SeqMoE achieves lower prefill latency through flexible prefetching and cache state transitions.

\begin{figure*}[t]
  \centering
  \includegraphics[width=\textwidth]{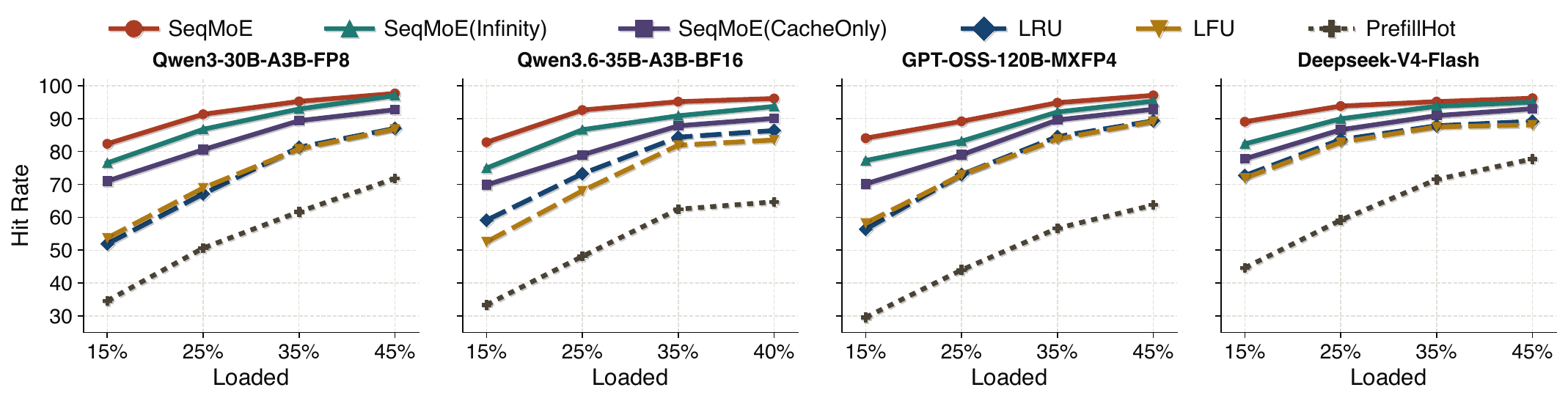}
  \caption{Expert Hit Rate Comparison.}
  \Description{...}
  \label{fig:ex_hit}
\end{figure*}

\begin{figure}[b]
  \centering
  \includegraphics[width=\linewidth]{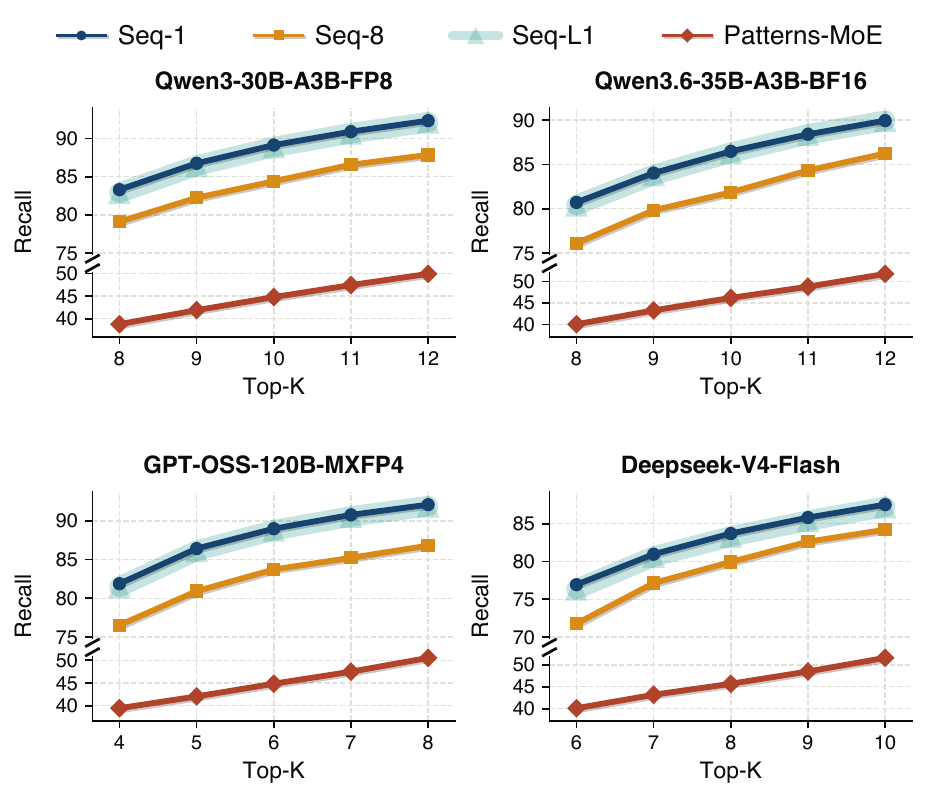}
  \caption{Expert Prediction Recall Comparison.}
  \Description{...}
  \label{fig:ex_accuracy}
\end{figure}

\subsection{Expert Hit Rates}
Fig.~\ref{fig:ex_hit} compares expert hit rates, a key indicator of memory-management efficiency and the potential to approach FullLoad performance. LRU, LFU, and PrefillHot are adopted by FreeToken, MoE-APEX, and KTransformers. PrefillHot selects the most frequently activated experts during prefill and keeps this selection fixed throughout decoding. At cache capacities of 15\%, 25\%, 35\%, 40\%, and 45\%, SeqMoE achieves average hit rates of 84.56\%, 91.72\%, 95.09\%, 96.11\%, and 96.97\%, respectively, compared with 60.02\%, 74.24\%, 84.49\%, 86.37\%, and 88.50\% for LRU, the strongest baseline.

We further evaluate two ablation variants: SeqMoE (Infinity), which uses score-based prefetch scheduling from MoE-Infinity, and SeqMoE (CacheOnly), which relies solely on forecast-driven caching. Future-aware eviction alone outperforms existing history-based policies, demonstrating the value of predicted activations in guiding cache residency. Prefetching further improves hit rates by loading experts before they are needed. However, independent score-based prioritization struggles to select the most beneficial combination of prefetch tasks under bandwidth and deadline constraints. This limitation is especially pronounced at smaller cache capacities, where higher miss rates generate more prefetch tasks and intensify bandwidth contention, highlighting the importance of joint deadline-aware scheduling.

\subsection{Prediction Accuracy}
\label{sec:Prediction Accuracy}
Fig.~\ref{fig:ex_accuracy} compares prediction recall. Patterns-MoE lacks sufficient accuracy for reliable prefetching, as fetching incorrect experts not only wastes transfer bandwidth but can also evict cached experts that would otherwise be reused. Seq-1 denotes one-step prediction over full sequences. Selecting the top $k+3$ experts yields recalls of 90.90\%, 88.39\%, 90.76\%, and 85.81\% on QW3, QW36, GPT, and DSV4, respectively. Seq-8 recursively predicts eight steps, with an average recall drop of only 3.58 percentage points relative to Seq-1. This limited degradation demonstrates robustness to recursive error accumulation and supports the effectiveness of activation-state preprocessing and scheduled sampling in maintaining accuracy over longer prediction horizons. Seq-L1 reports recall on sequences longer than 4,096 tokens and remains comparable to Seq-1, suggesting that prediction accuracy remains stable even for longer sequences within the evaluated range of 0.1K–10K tokens.

\subsection{Offloading Overhead}
\begin{figure}[t]
  \centering
  \includegraphics[width=\linewidth]{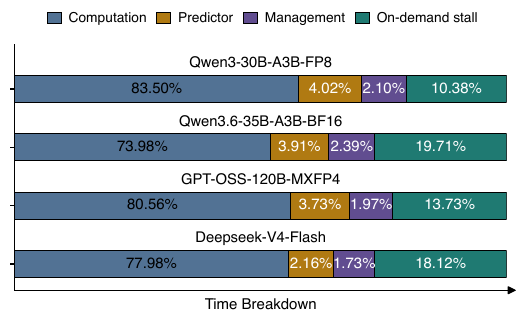}
  \caption{Inference Time Breakdown.}
  \Description{...}
  \label{fig:ex_breakdown}
\end{figure}

Fig.~\ref{fig:ex_breakdown} breaks down offloading inference time into four components at 45\% cache capacity (40\% for QW36). Offloading overhead comprises three components: on-demand stalls, predictor inference, and management operations. Despite an average miss rate of only 3.25\%, on-demand loading remains the largest component, highlighting the substantial cost of data transfers on the critical path and the priority of eliminating expert cache misses. Predictor inference accounts for 3.46\% of execution time on average. Increasing the number of predictors extends the overlap window but introduces additional prediction overhead and slightly reduces accuracy, creating a trade-off between lookahead, prediction cost, and reliability. The total predictor memory footprints are 120 MB, 228 MB, 112 MB, and 223 MB for QW3, QW36, GPT, and DSV4, respectively. Increasing the number of predictors does not increase their total memory footprint, since each predictor covers fewer layers and can therefore use a smaller hidden dimension. CUDA management kernels, including cache management, prefetch dispatch, and prefetch landing, account for 2.05\% of execution time on average.

\section{Conclusion}

We presented SeqMoE, an MoE offloading system that approaches full-load performance through predictive memory management and a graph-compatible runtime. By formulating expert activation prediction as sequence modeling, SeqMoE provides accurate, long-horizon forecasts that jointly guide prefetch scheduling and cache eviction. This coordinated design improves expert hit rates across the memory hierarchy to reduce on-demand stalls. Moreover, compute-transparent expert placement and synchronization-free orchestration disciplines reconcile dynamic offloading with graph-based execution mechanisms, enabling end-to-end graph capture and translating improved memory management into inference performance. Evaluations demonstrate that SeqMoE substantially outperforms existing offloading systems, highlighting the importance of jointly designing prediction, memory management, and execution to realize MoE’s structural advantages in high-performance, low-memory inference. By combining timely expert availability with efficient execution, SeqMoE advances the state of the art in MoE offloading inference.

\newpage
\bibliographystyle{ACM-Reference-Format}
\bibliography{ref.bib}

\end{document}